\documentclass[aps,10pt,pre,byrevtex,twocolumn,superscriptaddress,bibnotes,showpacs,showkeys,longbibliography]{revtex4-1}
\usepackage{microtype}
\usepackage{graphicx}
\usepackage{amsmath}
\usepackage{amssymb}
\usepackage[colorlinks,linkcolor=blue,urlcolor=blue,citecolor=blue]{hyperref}

\newcommand{\be}{\begin{equation}}
\newcommand{\ee}{\end{equation}}
\newcommand{\ra}{\rightarrow}
\newcommand{\cL}{\mathcal{L}}
\newcommand{\cH}{\mathcal{H}}
\newcommand{\reals}{\mathbb{R}}
\newcommand{\p}{\partial}
\newcommand{\tG}{\tilde G}

\begin{document}
\title{Large deviations for Markov processes with resetting}

\author{Janusz M. Meylahn}
\affiliation{Mathematical Institute, Leiden University, Leiden, The Netherlands}
\affiliation{\mbox{Institute of Theoretical Physics, Department of Physics, Stellenbosch University, Stellenbosch 7600, South Africa}}

\author{Sanjib Sabhapandit}
\affiliation{Raman Research Institute, Bangalore 560080, India}

\author{Hugo Touchette}
\email{htouchet@alum.mit.edu, htouchette@sun.ac.za}
\affiliation{National Institute for Theoretical Physics (NITheP), Stellenbosch 7600, South Africa}
\affiliation{\mbox{Institute of Theoretical Physics, Department of Physics, Stellenbosch University, Stellenbosch 7600, South Africa}}

\date{\today}

\begin{abstract}
Markov processes restarted or reset at random times to a fixed state or region in space have been actively studied recently in connection with random searches, foraging, and population dynamics. Here we study the large deviations of time-additive functions or observables of Markov processes with resetting. By deriving a renewal formula linking generating functions with and without resetting, we are able to obtain the rate function of such observables, characterizing the likelihood of their fluctuations in the long-time limit. We consider as an illustration the large deviations of the area of the Ornstein-Uhlenbeck process with resetting. Other applications involving diffusions, random walks, and jump processes with resetting or catastrophes are discussed.
\end{abstract}



\maketitle


\section{Introduction}

Stochastic processes with restarting or reset events,
corresponding to random transitions in time to a given state or region
in space, have been the subject of active studies in physics and
mathematics in recent years. In physics, such processes have been
studied as a mechanism for power-law distributions \cite{manrubia1999}
and, more recently, as random search models suggested by common
experience (e.g., losing one's keys) in which periods of
diffusive exploration are interspaced with random returns to a
starting point \cite{evans2011,evans2011b,evans2013,evans2014,kusmierz2014,gupta2014,majumdar2015b,majumdar2015c}.
In this context, a reset is also called a
restart \cite{janson2012} or a teleportation \cite{benichou2007} and
can be considered as part of more general intermittent search
strategies combining different exploration dynamics
\cite{benichou2011}.

In mathematics, processes with reset have been studied mostly in the
context of birth-death processes modelling the evolution of
populations in which partial or complete extinction or emigration
events happen at random times
\cite{pakes1978,brockwell1982,brockwell1985,kyriakidis1994,pakes1997,economou2003,dharmaraja2015}. In
this context, a reset is more often referred to as a catastrophe,
disaster or decimation and can also be seen as an absorbing or
``killing'' state that triggers, when reached, a restart or ``resurrection''
of the process \cite{pakes1997}. Similar jump processes have been
studied for modelling queues where random ``failures'' clearing the
content or occupation of a queue are followed by ``repaired phases'' in
which the queue functions normally \cite{crescenzo2003,crescenzo2012,kumar2000}.

The focus of these studies, both on the physical and mathematical sides, is on determining time-dependent and stationary distributions, as well as
survival and first-passage time statistics using modified Master or
Fokker-Planck equations that account for the effect of
resetting. Renewal representations of distributions and first-passage
time statistics have also been obtained for jump processes \cite{kyriakidis1994,pakes1997,economou2003}
and diffusion equations \cite{evans2011,evans2011b,evans2013}. First-passage times are especially important for search
applications, as they provide a measure of the efficiency of adding
resetting to random walks.

Here, we consider a different problem involving resetting, namely, that of deriving large
deviation functions for additive observables. The study of large deviations for ``normal''
Markov processes is an active area of probability theory having many applications in queueing theory,
estimation, and control
\cite{bucklew1990,shwartz1995,dembo1998}.  Large deviation functions
also play a fundamental role in statistical physics by providing
rigorous versions of the notions of entropy and free energy for
equilibrium systems \cite{ellis1985}, which can be generalized to
nonequilibrium systems driven in steady states
\cite{touchette2009,hollander2000,derrida2007,bertini2007,harris2013}. In
this context, an additive observable is simply a quantity integrated
over time for a physical system evolving stochastically due to the
influence of noise, external forces, and boundary reservoirs. It can
represent, for example, the work done when pulling a Brownian particle
with laser tweezers \cite{zon2003a}, the stretch of a molecular motor attached to a
protein \cite{meylahn2015}, or the total energy or particle current exchanged between
different reservoirs in a given time interval \cite{derrida2007}. In all cases, the
fluctuations of the observable studied are characterized in the
long-time limit by the so-called rate function, which is the
central function of large deviation theory \cite{ellis1985,dembo1998,hollander2000,touchette2009}.

We obtain in the following large deviation functions for processes with resetting by deriving two representations for the generating function of additive observables: one that is essentially a reset generalization of the Feynman-Kac formula and another that links, via a renewal argument, the generating function of an observable with resetting to its generating function without resetting. The derivation of rate functions follows from these results by studying, as is common in large deviation theory, the long-time asymptotics of generating functions. As an illustration of these results, we consider in Sec.~\ref{secex} the large deviations of the integral (area) of the reset Ornstein-Uhlenbeck process, which can be considered as a simple model of molecular motor with resetting \cite{meylahn2015}. Other applications related to birth-death processes and queues are mentioned in the conclusion of the paper.

\section{Problem}
\label{secmod}

To simplify the presentation, we consider the case of one-dimensional diffusions. Higher-dimensional diffusions and jump processes such as birth-death processes can be treated with minor changes of notation.

We thus consider an ergodic diffusion process $X_t\in\reals$ described by the stochastic differential equation (SDE)
\be
dX_t=F(X_t)dt +\sigma dW_t,
\label{eqsde1}
\ee
which is reset to the fixed position $x_r$ at random times distributed according to an exponential distribution with parameter $r\geq 0$. Considering the evolution of $X_t$ over an infinitesimal time $dt$, this means that $X_t$ is either reset to $X_{t+dt}=x_r$ with probability $rdt$ or that $X_t$ diffuses with probability $1-rdt$ according to the SDE (\ref{eqsde1}), which involves the drift $F(X_t)$, the noise power $\sigma>0$, and the Brownian motion or Wiener process $W_t$. 

As shown in \cite{evans2011,evans2011b}, the resetting modifies the Fokker-Planck equation governing the evolution of the probability density $p(x,t|x_0)$ of $X_t$ started at $X_0=x_0$ by adding a uniform sink and a source at $x_r$:
\begin{eqnarray}
\frac{\p}{\p t}p(x,t|x_0)&=&-\frac{\p}{\p x} F(x) p(x,t|x_0)+\frac{\sigma^2}{2}\frac{\p^2}{\p x^2}p(x,t|x_0)\nonumber\\
& & \qquad -rp(x,t|x_0)+r\delta(x-x_r).
\label{eqfp1}
\end{eqnarray}
Alternatively, $p(x,t|x_0)$ can be obtained by noting that $X_t$ can reach $x\neq x_r$ by diffusing either from its last reset position $x_r$, which occurred at the random time $t-\tau$, or from its initial state $x_0$ without resetting, so that
\be
p(x,t|x_0)=e^{-rt} p_0(x,t|x_0)+\int_0^t r e^{-r\tau}\, p_0(x,\tau|x_r) d\tau,
\label{eqrenew1}
\ee
where $p_0(x,t|x_0)$ is the free propagator solving the standard Fokker-Planck equation (\ref{eqfp1}) with $r=0$ \cite{evans2011,evans2011b}. Similar renewal formulae connecting time-dependent distributions with and without resetting have been obtained in the context of jump processes modelling population dynamics \cite{brockwell1982,brockwell1985,kyriakidis1994,economou2003} and queues \cite{kumar2000,crescenzo2003,crescenzo2012}. Modified Fokker-Planck equations with resetting have also been obtained by studying the diffusive or Kramers-Moyal limit of reset jump processes; see \cite{crescenzo2003,crescenzo2012,dharmaraja2015,durang2014}. 

Here, we  study the probability density not of the process itself but of functionals or observables of $X_t$ having the general time-additive form
\be
A_T=\frac{1}{T}\int_0^T f(X_t)dt,
\label{eq:addobs}
\ee where $f$ is a real function of $X_t$. Such observables naturally
arise in manmade and physical systems, as mentioned, and are often characterized by
a probability density having the form
\be 
P(A_T=a)= e^{-T I(a)+o(T)} 
\label{LDform}
\ee 
in the limit of large integration times $T$, with $o(T)$ denoting any
correction term that grows slower than $T$. This scaling of
probabilities is known in large deviation theory as the
\emph{large deviation principle} (LDP)
\cite{ellis1985,dembo1998,hollander2000,touchette2009} and implies
that fluctuations of $A_T$ are exponentially unlikely to be observed in
the long-time limit. This applies for all values $A_T=a$ such that the
rate of decay or rate function $I(a)$ is positive. In general, $I(a)$ also has (at least) one zero
$a^*$ determining the \emph{typical value} of $A_T$ around which
$P(A_T=a)$ concentrates exponentially as $T\ra\infty$. The rate
function is thus important as it characterizes in the long-time limit
the typical value of $A_T$, which corresponds to its ergodic or
stationary value, as well as the atypical fluctuations around this
ergodic value.

For processes with no resetting, the rate function is generally obtained by calculating the \emph{scaled cumulant generating function} (SCGF) of $A_T$ defined by the limit
\be
\lambda_0(k)=\lim_{T\ra\infty} \frac{1}{T}\ln E_{x}^0\big[e^{TkA_T}\big],
\label{eqscgf1}
\ee
where $k\in\reals$ and $E_{x}^0[\cdot]$ denotes the expectation with respect to the non-reset process $X_t$ started at $X_0=x$. For Markov processes, it is known that this function coincides under general conditions with the dominant eigenvalue of the so-called \emph{tilted generator} \cite{ellis1985,dembo1998,hollander2000,touchette2009}, which for the SDE (\ref{eqsde1}) has the form
\be
\cL_k=L+kf,
\label{eqtgen1}
\ee
where
\be
L=F\frac{\p}{\p x}+\frac{\sigma^2}{2}\frac{\p^2}{\p x^2}
\label{eqgen1}
\ee
is the generator of the diffusion $X_t$ without resetting. In this case, the calculation of large deviations is therefore essentially a spectral problem. Assuming that $\lambda_0(k)$ can be obtained and is differentiable, we then have from an important result of large deviation theory, known as the G\"artner-Ellis Theorem \cite{ellis1985,dembo1998,hollander2000,touchette2009}, that $A_T$ satisfies an LDP with rate function $I_0(a)$ given by the Legendre-Fenchel transform of the SCGF:
\be
I_0(a)=\sup_{k}\big\{ka-\lambda_0(k)\big\}.
\label{eqge1}
\ee
 
This method can be applied in principle to processes with resetting, but the generator of $X_t$ in this case is not a pure differential operator: it is a mixed operator involving the pure part (\ref{eqgen1}) and a singular integral kernel accounting for the delta source in the Fokker-Planck equation. Finding the SCGF by spectral method then becomes a complicated and singular problem, so that other methods must be used. We propose one in the next section based on the renewal representation of reset processes.

\section{Results}
\label{secres}

We obtain the large deviations of $A_T$ for the process $X_t$ with resetting by studying, following the limit (\ref{eqscgf1}), the time evolution of the generating function:
\be
G_r(x,k,t)=E_{x}\big[e^{tkA_t}\big]=E_{x}\big[e^{k\int_0^tf(X_s)ds}\big],
\ee
where $E_x[\cdot]$ denotes the expectation with respect to the process $X_t$ with resetting started at $X_0=x$. Without resetting ($r=0$), this function is known to evolve according to the Feynman-Kac (FK) formula
\be
\frac{\p}{\p t} G_0=\cL_k G_0,
\label{eqfk1}
\ee
which is a parabolic linear partial differential equation for $G_0=G_{r=0}$ with initial condition $G_0(x,k,0)=1$ \cite{majumdar2005}. 

A modified FK formula that includes resetting can be derived similarly to the reset-free case by considering an additional time step $dt$ in the generating function, so as to write
\begin{eqnarray}
G_r(x,k,t+dt) &=&E_x\big[e^{\int_0^{dt}f(X_s)ds}e^{\int_{dt}^{t+dt}f(X_s)ds}\big]\nonumber\\
&=& e^{f(x)dt} E_x\big[e^{\int_{dt}^{t+dt}f(X_s)ds}\big],
\end{eqnarray}
using $X_0=x$. From this initial state, the process can either reset to $X_{dt}=x_r$ with probability $rdt$ or diffuse to $X_{dt}$ according to the SDE (\ref{eqsde1}) with the complementary probability $1-rdt$, so that
\begin{multline}
G_r(x,k,t+dt)=e^{f(x)dt}\bigg\{rdt\, G_{r}(x_r,k,t)\bigg. \\ 
\bigg. +(1-rdt) \int_{-\infty}^\infty d\xi\, K(\xi) G_r(x+\xi,k,t) \bigg\},
\end{multline}
where $K(\xi)$ is the probability distribution of the increment $X_{dt}-X_0=\xi$ as determined from (\ref{eqsde1}). In this way, we separate the resetting from the pure diffusion (\ref{eqsde1}). Expanding $G_r(x+\xi,k,t)$ up to second order in $\xi$ and performing the integral then yields
\be
\frac{\p}{\p t}G_r=(\cL_k -r)G_r+rG_r(x_r,k,t)
\label{eq:FKev}
\ee
with the initial condition $G_r(x,k,0)=1$.

This modified FK formula with uniform sink and source at $x_r$ is similar to equations obtained for the first-passage problem with resetting \cite{evans2011,evans2011b,evans2013} and must be solved, as for this problem, by considering the source term $G_r(x_r,k,t)$  as a constant and by matching the solution $G_r(x,k,t)$ self-consistently for $x=x_r$. This is a difficult task in general, which does not suggest in our experience an efficient way to obtain large deviations, especially since we need the generating function for large times in order to obtain the limit
\be
\lambda_r(k)=\lim_{T\ra\infty} \frac{1}{T}\ln G_r(x,k,T),
\label{eqscgf2}
\ee
which is the reset version of (\ref{eqscgf1}).

For the purpose of calculating this limit, a more useful renewal representation of $G_r(x,k,t)$ similar to (\ref{eqrenew1}) can be derived. To this end, assume that the time interval $[0,T]$ witnesses $n$ resettings with periods $\tau_1,\tau_2,\ldots,\tau_{n}$ such that
\be
T=\sum_{i=1}^{n+1} \tau_i
\label{eqtimecons1}
\ee
and
\be
TA_{T} = \sum_{i=1}^{n+1}\int_{\sum_{j=1}^{i}\tau_{j-1}}^{\tau_{i}}f(X_{s})ds,
\ee
where $\tau_{n+1}$ is the last period without resetting leading to $T$. Because of the additive form of $A_T$, it is clear that $G_r$ can be decomposed, when conditioned on these $n$ resettings, into a product of generating functions $G_0$ involving only pure diffusion between resettings. To write the full $G_r$, we then have to sum over all possible reset number and reset times. Since the probability of having a reset at time $\tau$ is $re^{-r\tau}$ and the probability of no reset until the time $\tau$ is $e^{-r\tau}$, we thus obtain 
\begin{widetext}
\be
G_{r}(x, k, T) = \sum_{n=0}^{\infty}\int_0^T d\tau_1\, re^{-r\tau_1}G_{0}(x, k, \tau_1)
\int_0^Td\tau_2\, re^{-r\tau_2}G_0(x_r, k,\tau_2)\cdots
\int_0^Td\tau_{n+1}\, e^{-r\tau_{n+1}}G_0(x_r, k, \tau_{n+1})
\, \delta\bigg(T - \sum_{i=1}^{n+1}\tau_{i}\bigg).
\label{eq:gensplit}
\ee
\end{widetext}
Notice that the first $G_0$ term starts at the initial condition $X_0=x$, while the others start after resetting at $x_r$. The probability of the last period $\tau_{n+1}$ is also different from the other periods, since it is determined by the prior $n$ reset periods and the constraint (\ref{eqtimecons1}), included in (\ref{eq:gensplit}) with the delta function.

To deal with this constraint, it is natural to consider the Laplace transform in time of the generating function
\be
\tG_r(x,k,s)=\int_0^\infty G_r(x,k,T)e^{-sT}\, dT,
\ee
which yields, after integration over the $\tau_i$'s,
\be
\tG_r(x,k,s)=\tG_0(x,k,s+r)\sum_{n=0}^\infty r^n \tG_0(x_r,k,s+r)^n,
\ee
where $\tG_0$ denotes the Laplace transform of $G_0$. Assuming that
\be
r\tG_0(x_r,k,s+r)<1,
\ee
we therefore obtain
\be
\tG_r(x, k, s) = \frac{\tG_0(x,k,s+r)}{1 - r \tG_0(x_r,k,s+r)}.
\label{eq:finalRR}
\ee

This is our main result connecting the generating function of $A_T$ with resetting to its generating function without resetting. It can be verified that this formula is equivalent (by Laplace transform) to the modified FK equation \eqref{eq:FKev}, though \eqref{eq:finalRR} is simpler, as it expresses $G_r$ explicitly in terms of the free generating function $G_0$. 

This is more convenient for obtaining large deviations. Assuming that the limit (\ref{eqscgf2}) defining the SCGF $\lambda_r(k)$ of $A_T$ exists implies the following scaling of the generating function:
\be
G_r(x,k,T)\sim e^{\lambda_r(k)T}
\ee
as $T\ra\infty$, which translates in Laplace space into
\be
\tG_r(x,k,s)\sim\frac{1}{s-\lambda_r(k)}.
\label{eqpole1}
\ee
As a result, we see that the SCGF of $A_T$ for the resetting process can be determined by locating the largest (simple and real) pole of the right-hand side of \eqref{eq:finalRR}, which is also a zero (in $s$) of the denominator $1-r\tG_0$ when $\tG_0$ is finite. If $\lambda_r(k)$ is differentiable as a function of $k$, we then obtain the rate function $I_r(a)$ of $A_T$ similarly to (\ref{eqge1}) by taking the Legendre-Fenchel transform of $\lambda_r(k)$.

These calculations are based only on the knowledge of the generating function $G_0$ of $A_T$ without resetting. In some cases, the large-time asymptotics of that generating function proves to be sufficient to obtain the desired pole $\lambda_r(k)$, which means that the large deviations of $A_T$ for the process with resetting can be obtained directly from the large deviations of $A_T$ without resetting. This important result is illustrated next.

\section{Example}
\label{secex}

We consider in this section the reset Langevin equation (or reset Ornstein-Uhlenbeck process) obtained by adding resettings at $x_r$ with rate $r$ to the diffusion
\begin{equation}
dX_t = -\gamma X_t dt + \sigma dW_t,
\label{eq:OUprocess}
\end{equation}
where $\gamma$ is the friction coefficient, $\sigma$ is the noise strength, and $W_t$ is the Wiener process. The
stationary distribution of this model was studied recently in~\cite{pal2015}.
The observable that we consider is the integral of the state,
\be
A_T=\frac{1}{T}\int_0^T X_t dt.
\label{eqarea1}
\ee

This reset process can be considered physically as a simple model of filament dynamics in motility assays \cite{schaller2010,schaller2011,sumino2012}, wherein filaments are pulled by spring-like motor proteins attached to a substrate at one end and moving on filaments at the other \cite{banerjee2011}. In this context, $A_T$ represents the mean force exerted on one filament over a time $T$, which is proportional to the stretch $X_t$ of the motor protein attached to it, while resetting happens when the motor randomly detaches from the filament and a new motor attaches itself with zero stretch \cite{meylahn2015}.

The generating function $G_0$ of $A_T$ for the reset-free Langevin equation is known in closed form, but its Laplace transform is relatively complicated to work with. For our purpose, it is more convenient to expand $G_0$, following the FK formula (\ref{eqfk1}), in spectral form as
\be
G_0(x, k, T) = \sum_{i=0}^{\infty} \psi_{k, i}(x)e^{\lambda_{0,i}(k)T},
\label{eq:genfuncexpansion}
\ee
where $\lambda_{0,i}(k)$ are the eigenvalues of the tilted generator $\cL_k$ without resetting and $\psi_{k, i}$ are the corresponding eigenfunctions. Such a spectral decomposition can be obtained in principle for any Markov process. By symmetrization to the quantum oscillator (see the Appendix), we explicitly find here
\be
\lambda_{0,i}(k)=\frac{k^2\sigma^2}{2\gamma^2}-i\gamma,\quad i=0,1,\ldots 
\label{eqeigval1}
\ee
and
\be
\psi_{k,i}(x)=\frac{(-1)^i \gamma ^{-3 i/2} k^i \sigma ^i e^{\frac{k x}{\gamma }-\frac{3 k^2 \sigma ^2}{4 \gamma ^3}}
   H_i\left(\frac{\sqrt{\gamma } x}{\sigma }-\frac{k \sigma }{\gamma ^{3/2}}\right)}{\sqrt{2^i i!} \sqrt{(2
   i){!!}}}
\ee
where $H_i$ is $i$th Hermite polynomial. The SCGF $\lambda_0(k)$ of $A_T$ corresponds to the largest eigenvalue:
\be
\lambda_0(k)=\max_i \lambda_{0,i}(k)=\frac{\sigma^2k^2}{2\gamma^2}.
\ee
From the Legendre-Fenchel transform (\ref{eqge1}), we thus find the rate function of $A_T$ without resetting to be
\be
I_0(a)=\frac{\gamma^2 a^2}{2\sigma^2},
\label{eqgauss1}
\ee
which implies that the fluctuations of $A_T$ are Gaussian-distributed around $A_T=0$ \footnote{This is also evident from the fact that $A_T$ is a linear integral of a linear Gaussian process.}.

\begin{figure}[t]
\centering
\resizebox{3in}{!}{\includegraphics{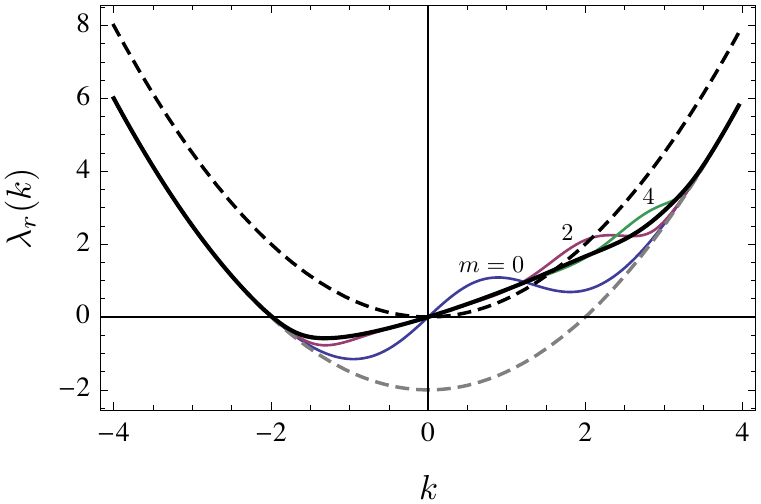}}
\caption{(Color online) Dominant pole of $\tG_r(x,k,s)$ for increasing truncation orders: $m=0$ (blue), $m=2$ (purple), $m=4$ (green). Black curve: Convex $\lambda_r(k)$ obtained for $m\geq 6$. Dashed black curve: Non-reset $\lambda_0(k)$. Dashed gray curve: Tail approximation of $\lambda_r(k)$ shown in (\ref{eqdef2}). Parameters: $x_r=1$, $r=2$, $\gamma=1$, $\sigma=1$.}
\label{figscgf1}
\end{figure}

To determine the effect of resetting on these fluctuations, we insert the Laplace transform of the spectral representation (\ref{eq:genfuncexpansion}),
\be
\tG_0(x,k,s)=\sum_{i=0}^\infty \frac{\psi_{k,i}(x)}{s-\lambda_{0,i}(k)},
\ee
into the Laplace formula (\ref{eq:finalRR}) and locate the largest pole of the resulting expression for a given truncation $0\leq i\leq m$. The result is shown for $x_r=1$, $r=2$, and various truncation orders $m$ in Fig.~\ref{figscgf1} and compared with the reset-free SCGF $\lambda_0(k)$. As can be seen, the dominant pole is nonconvex in $k$ for low truncation orders, which means that it does not represent a valid SCGF, since SCGFs are always convex by definition \cite{ellis1985,dembo1998,hollander2000,touchette2009}. By increasing however the truncation order, the pole does converge to a convex function, identified from (\ref{eqpole1}) as $\lambda_r(k)$. For the parameter values used in Fig.~\ref{figscgf1}, convergence is attained essentially for $m\gtrsim 6$; for larger values of $|x_r|$ or $r$, more modes are generally required. 

This applies to the part of $\lambda_r(k)$ close to $k=0$, which describes the small fluctuations of $A_T$. For the large fluctuations associated with the tails of $\lambda_r(k)$, convergence appears immediately for one mode, as can be seen in Fig.~\ref{figscgf1}, which implies the following approximation: 
\be
\lambda_r(k)\approx\lambda_0(k)-r+r \psi_{k,0}(x_r).
\label{eqdef1}
\ee
Here, we have explicitly
\be
\psi_{k,0}(x)=e^{k x/\gamma-3 k^2 \sigma ^2/(4 \gamma^3)},
\ee
so that (\ref{eqdef1}) can be simplified in fact to
\be
\lambda_r(k)\approx\lambda_0(k)-r
\label{eqdef2}
\ee
for $|k|\ra\infty$.

This simple tail behavior of $\lambda_r(k)$ can be understood by noting that very large fluctuations of $A_T$ are brought about, for relatively small reset positions $x_r$, by long excursions of the process far away from $x_r$ having very few or no reset events. As a result, the renewal representation (\ref{eq:gensplit}) is dominated by purely diffusive trajectories whose large deviations are determined by the dominant mode of $G_0$ as $T\ra\infty$. The $r$ factor in (\ref{eqdef2}) only accounts for the probability of seeing such trajectories without resetting. Conversely, more modes of $G_0$ are needed to describe the small fluctuations of $A_T$ close to $x_r$ because such fluctuations are brought about by trajectories that have many resettings and, therefore, many short diffusive trajectories for which the large deviation limit is not effective. The number of modes $m$ that must be used to recover the correct $\lambda_r(k)$ depends on the parameters used: generally, the larger $r$ or $|x_r|$ is, the higher $m$ should be since resetting takes place more often.

\begin{figure}[t]
\centering
\resizebox{3in}{!}{\includegraphics{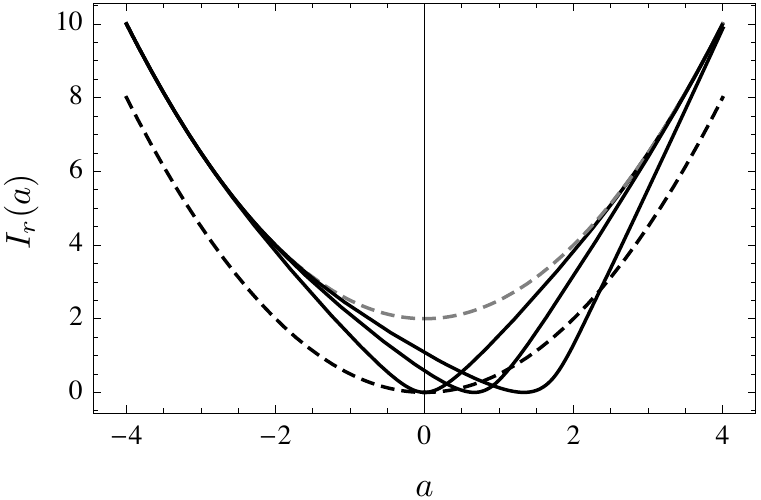}}
\caption{Black curves: $I_r(a)$ for $x_r =0,1,2$ (from left to right). The first two curves were obtained for $m=10$, while the last for $x_r=2$ was obtained for $m=20$. Dashed black curve: Non-reset rate function $I_0(a)$. Dashed gray curve: Tail approximation of $I_r(a)$ shown in (\ref{eqiapprox1}). Parameters: $r=2$, $\gamma=1$, $\sigma=1$.}
\label{figratef1}
\end{figure}

\begin{figure}[t]
\centering
\resizebox{3in}{!}{\includegraphics{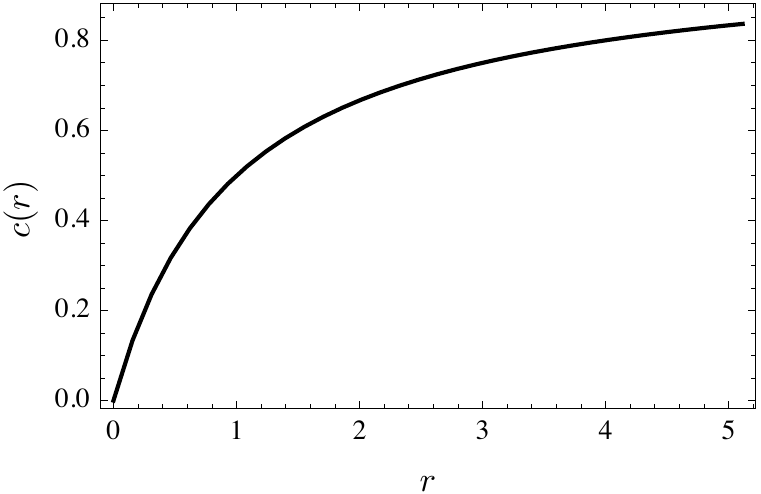}}
\caption{Proportionality coefficient $c(r)$ between the minimum $a^*$ of $I_r(a)$ and the resetting position $x_r$. Parameters: $\gamma=1$, $\sigma=1$.}
\label{figcoef1}
\end{figure}

Once that number is set, the rate function $I_r(a)$ can be computed as the Legendre-Fenchel transform of $\lambda_r(k)$. The result is shown in Fig.~\ref{figratef1} for $r=2$ and different resetting positions $x_r$. As expected, the rate function $I_r$ is narrower than $I_0$ and shifts towards the resetting position $x_r$, since $X_t$ is more likely with resetting to stay near $x_r$. Note, however, that the minimum $a^*$ of the rate function, corresponding to the most probable value of $A_T$ in the ergodic limit $T\ra\infty$, is not exactly $x_r$ because the friction in the Langevin equation brings $X_t$ near $x=0$. It is difficult to study this competing effect analytically, since it is strongly linked to resetting, and so cannot be treated perturbatively using a mode expansion of $G_0$. Numerically, we find that $a^*$ varies linearly with $x_r$ with a slope $c(r)$ shown in Fig.~\ref{figcoef1}. As $r\ra\infty$, $c(r)\ra1$, and thus $a^*\ra x_r$, as expected.

Looking back at Fig.~\ref{figratef1}, we can also see that the tails of $I_r(a)$ are mostly unaffected by resetting, except for a constant shift. This comes again from the large fluctuations of $A_T$ being the result of large diffusive excursions that have very few or no resetting events, so that (\ref{eqdef2}) holds. Inserting this tail approximation into the Legendre-Fenchel transform leads to the dual approximation
\be
I_r(a)\approx I_0(a)+r
\label{eqiapprox1}
\ee
as $|a|\ra\infty$. This gives a good approximation of the rate function, as can be seen in Fig.~\ref{figratef1}.

This tail result implies with (\ref{eqgauss1}) that $A_T$ has large Gaussian fluctuations, reflecting with a shift its Gaussian fluctuations (\ref{eqgauss1}) seen without resetting. The small fluctuations of $A_T$ around its typical value and mean $a^*$ are also Gaussian, as can be seen by expanding $I_r(a)$ around its minimum $a^*$, but with a reset-modified variance determined by $\lambda_r''(0)$ or $I''_r(a^*)^{-1}$ \cite{touchette2009}. Finally, in the intermediate region away from $a^*$, where (\ref{eqiapprox1}) is not an accurate approximation of $I_r(a)$, the competition between resetting and diffusion leads to non-Gaussian fluctuations, characterized by the non-parabolic rate function seen in Fig.~\ref{figratef1}.

\section{Conclusion}
\label{secconc}

We have derived in this paper a general renewal formula (\ref{eq:finalRR}) that can be used to obtain the large deviation functions of additive observables of Markov processes with resetting, and have illustrated this result for the Langevin equation with resetting. Other applications should follow this example either via the exact calculation of the generating function or via the general spectral expansion (\ref{eq:genfuncexpansion}), keeping in mind for this expansion to include enough modes, as demonstrated, to obtain properly scaled convex cumulant generating functions in the long-time limit.

Although we have considered reset diffusions, it is clear that our main results expressed in terms of generating functions also hold for birth-death and jump processes in general, in addition to Markov chains with resetting or catastrophes, thus opening the way for many other applications. In birth-death processes, for example, one could consider as an observable the total number of births over a given time period or any birth-related cost (e.g., insurances) accumulated in that period. Similarly, for queueing models with resetting, the observable may represent the number of clients entering a queue or any cost associated with clients which is additive in time. 

For these examples, we expect the main results that we have obtained for the reset Langevin equation to hold. In particular, it is clear that as long as large fluctuations of $A_T$ are the results of long trajectories involving few resetting, as is the case for the Langevin equation, then the large deviations functions obtained with reset are a shift of the large deviations obtained without reset, following the approximations (\ref{eqdef2}) and (\ref{eqiapprox1}) that we have derived, with the shift coming from the probability of having few or no resettings over the time $T$.  

For future work, it would be interesting to study whether observables that do not have a large deviation principle without resetting acquire that principle when resetting is introduced. It is known that resetting adds an effective confinement that can transform a non-stationary process (e.g., Brownian motion \cite{evans2011}) into a stationary one, but this might not be enough on its own to force a large deviation principle. Another interesting problem is to generalize our results to observables involving an integral of the increments of the process considered (in the case of pure diffusions) or a sum over its jumps (in the case of pure jump processes); see \cite{chetrite2014} for more detail. These observables represent physically quantities, such as particle currents and entropy production, playing an important role in nonequilibrium statistical physics.

\appendix*
\section{Spectral decomposition of $G_0(x,k,T)$}
\label{appsym}

The generating function $G_0(x,k,T)$ evolves without resetting according to the linear partial differential (\ref{eqfk1}) and can therefore be decomposed in the eigenbasis of the tilted generator $\cL_k$, shown in (\ref{eqtgen1}). For the Ornstein-Uhlenbeck process, $\cL_k$ is not hermitian, but can be mapped via a unitary transformation to a hermitian, Schr\"odinger-type operator, so its spectrum is real. This transformation or symmetrization is the same as the one used for the Fokker-Planck equation; see, e.g., Sec.~5.4 of \cite{risken1996}. 

Denote by $\rho(x)=e^{-U(x)}$ the stationary distribution of $X_t$ satisfying $L^\dagger \rho=0$. The symmetrization of $\cL_k$ is given by 
\be
\cH_k=\rho^{1/2}\cL_k\rho^{-1/2}=e^{-U/2}\cL_ke^{U/2}.
\ee
For the Ornstein-Uhlenbeck process, we have $U(x)=\gamma x^2/\sigma^2$ up to a constant, which leads to
\be
\cH_k=\frac{\sigma^2}{2}\frac{d^2}{dx^2}-\frac{\gamma^2 x^2}{2\sigma^2}+\frac{\gamma}{2}+kx.
\ee
This is the Schr\"odinger operator of a shifted and inverted quantum harmonic oscillator with mass $m=1$ and $\hbar=\sigma$ \cite{majumdar2002}. From the known spectrum of the harmonic oscillator, we therefore arrive at the eigenvalues (\ref{eqeigval1}). As for the eigenfunctions $\psi_{k,i}$, they are obtained by
\be
\psi_{k,i}(x)=\rho(x)^{-1/2}\Psi_{k,i}(x)=e^{U(x)/2}\Psi_{k,i}(x),
\ee
where $\Psi_{k,i}$ are the eigenfunctions of $\cH_k$, normalized in the usual quantum way.

\begin{acknowledgments}
J.M.M.\ was supported by bursaries from the National Institute for Theoretical Physics (NITheP) and the Harry Crossley Foundation. The work of H.T.\ was supported by the  National Research Foundation of South Africa (Grants no.\ 90322 and no.\ 96199) and Stellenbosch University (Project Funding for New Appointee).
\end{acknowledgments}

\bibliography{masterbib}

\end{document}